\documentclass{llncs}

\usepackage{graphicx}
\usepackage{wrapfig}
\usepackage{latexsym}

\newcommand{\outputs}{\mathsf{outputs}}
\newcommand{\sset}[2]{\left\{~#1  \left|
      \begin{array}{l}#2\end{array}
    \right.     \right\}}

\newcommand{\algorithm}[1]{\mbox{\sc\tt{#1}}}

\begin{document}

\title{Optimizing a Certified Proof Checker for a Large-Scale Computer-Generated Proof}
\author{Lu\'{i}s Cruz-Filipe \and Peter Schneider-Kamp}
\authorrunning{L.~Cruz-Filipe and P.~Schneider-Kamp}
\institute{Dept.\ Mathematics and Computer Science,
  Univ.\ Southern Denmark\\
  Campusvej 55, 5230 ODENSE M, Denmark\\
  \email{$\{$lcf,petersk$\}$@imada.sdu.dk}
}

\maketitle

\begin{abstract}
In recent work, we formalized the theory of optimal-size sorting networks with the goal of extracting a verified checker for the large-scale computer-generated proof that $25$~comparisons are optimal when sorting $9$~inputs, which required more than a decade of CPU time and produced $27$~GB of proof witnesses.
The checker uses an untrusted oracle based on these witnesses and is able to verify the smaller case of $8$~inputs within a couple of days, but it did not scale to the full proof for $9$ inputs.

In this paper, we describe several non-trivial optimizations of the algorithm in the checker, obtained by appropriately changing the formalization and capitalizing on the symbiosis with an adequate implementation of the oracle. We provide experimental evidence of orders of magnitude improvements to both runtime and memory footprint for $8$~inputs, and actually manage
to check the full proof for $9$~inputs.
\end{abstract}

\section{Introduction}
Sorting networks are hardware-oriented algorithms to sort a fixed number of inputs using a predetermined sequence of comparisons between them.
They are built from a primitive operator -- the \emph{comparator} --, which reads the values on two channels, and interchanges them if necessary to guarantee that the smallest one is always on a predetermined channel.
Comparisons between independent pairs of values can be performed in parallel, and the two main optimization problems one wants to addressed are: how many comparators do we need to sort $n$ inputs (the \emph{optimal size} problem); and how many computation steps do we need to sort $n$ inputs (the \emph{optimal depth} problem).

In previous work~\cite{ourICTAIpaper}, we proposed a generate-and-prune algorithm to show size optimality of sorting networks, and used it to show that $25$-comparator sorting networks have optimal size for $9$ inputs. The proof was performed on a massively parallel cluster and consumed more than $10$ years of computational time.
During execution we recorded the results of \emph{successful} search routines that allowed for reduction of the search space,
resulting in approx.\ $27$ GB of witnesses. 

Subsequently~\cite{ourformalization}, we formalized the relevant theory of sorting networks in Coq, therefrom extracting a certified checker able to confirm the validity of our informal computer-generated proof.
The checker bypasses the original search steps by means of an untrusted oracle, implemented by reading the log file produced by the original program,
and %
could verify the proof for the smaller case of $8$ inputs, thereby constituting the first computer-validated proof of the results in~\cite{Knuth66}.
However, due to the much larger dimension of the oracle, verifying the full proof for $9$ inputs was estimated to require approx.\ $20$ years of (non-parallelizable) computation.

In this paper, we show how careful optimizations of the formalization result in runtime improvement of several orders of magnitude, as well as drastic reductions of the memory footprint for the checker.
Throughout the paper, we benchmark the impact of the individual improvements using the feasible case of $8$ inputs, until we are able to check the full proof for $9$ inputs using around one week of computation on a Intel Xeon E5 clocked at $2.4$~GHz with $64$~GB of RAM.

Section~\ref{sec:background} shortly introduces the basic of sorting networks, the generate-and-prune algorithm from \cite{ourICTAIpaper}, and our formalization from \cite{ourformalization} to the degree necessary to understand the improvements. In Section~\ref{sec:runtime1} we change the checker algorithm in the formalization in order to bring runtime down by at least an order of magnitude, while we reduce memory footprint by a factor of $3$ in Section~\ref{sec:memory1}. Further substantial improvements to runtime and memory footprint are described in Sections~\ref{sec:runtime2} and \ref{sec:memory2}, respectively. We conclude in Sections~\ref{sec:conclusion} with a summary of the results and an outlook to possible future work.

\subsection{Related work}
The Curry--Howard correspondence states that every constructive proof of an existential statement embodies an algorithm to produce a witness of the required property.
This correspondence has been made more precise by the development of program extraction mechanisms for the most popular theorem provers.
In this paper, we focus on extracting a program from a Coq formalization, using the mechanism described in~\cite{DBLP:conf/cie/Letouzey08}.

Early experiments of program extraction from a large-scale formalization that was built form a purely mathematical perspective showed however that it is unreasonable to expect \emph{efficient} program extraction as a side result of formalizing textbook proofs~\cite{lcf:let:05}.
In spite of that, one can actually develop mathematically-minded formalizations that yield efficient extracted programs with only minor attention to definitions~\cite{DBLP:conf/tphol/OConnor08,DBLP:conf/mkm/KrebbersS11}.
This is in contrast with formalizations built with extraction as a primary goal, such as those in the CompCert project~\cite{DBLP:journals/cacm/Leroy09},
or with strategies that potentially compromise the validity of the extracted program (e.g.~using imperative data structures as in~\cite{DBLP:conf/tlca/Oury03}).

In this work we go one step further, and show that if the extracted program does not perform well enough, we can optimize it by tweaking the formalization without significantly changing it. The latter means less work reproving lemmas and theorems and ensures that the formalization remains understandable, in turn giving us confidence that we actually prove what we wish to prove.

Our contributions rely on the idea of an \emph{untrusted oracle}~\cite{DBLP:conf/sas/FouilheMP13,DBLP:journals/cacm/Leroy09}, where the extracted program checks the result of computations obtained through the oracle.
More specifically, we use an \emph{offline untrusted oracle}, where computation and checking are separated by logging the results of computations to a file.
This separation allows the use of massively parallel clusters for computation and the cheap reuse of the results during the development of the formalization and the checker.
In particular, we capitalize on the ability to pre-process the computational results offline to optimize the checker.

This offline approach to untrusted oracles is found in work on termination proofs~\cite{DBLP:conf/itp/Thiemann13,DBLP:conf/rta/ContejeanCFPU11}, where the separation is necessary as informal proof tools and checkers are modular programs developed by different research units. The difference to our work is the scale of the proofs: typical termination proofs have $10$-$100$ proof witnesses and total at most a few MB of data.
Recent work mentions that problems were encountered when considering proofs of ``several hundred megabytes'' \cite{DBLP:journals/corr/SternagelT14}.
In contrast, verifying the proof of size-optimality of sorting networks with $9$ inputs uses nearly $70$ million proof witnesses, totalling $27$ GB of oracle data.\\[-4.5ex]

\section{Background}
\label{sec:background}
We briefly summarize the key notions relevant to this work.
The interested reader is referred to~\cite{Knuth73} for a more extensive introduction to sorting networks, and to~\cite{ourICTAIpaper} for a detailed description of the proof we verify.

A \emph{comparator network} $C$ with $n$ channels and size $k$ is a
sequence of \emph{comparators} $C = (i_1,j_1);\ldots;(i_k,j_k)$, where
each comparator $(i_\ell,j_\ell)$ is a pair of channels $1\leq i_\ell< j_\ell\leq n$.
If $C_1$ and $C_2$ are comparator networks with
$n$~channels, then
$C_1;C_2$ denotes the comparator network obtained by concatenating
$C_1$ and $C_2$.
An input $\vec x=x_1\ldots x_n\in\{0,1\}^n$ propagates through $C$ as
follows: $\vec x^0 = \vec x$, and for $0<\ell\leq k$, $\vec x^\ell$ is
the permutation of $\vec x^{\ell-1}$ obtained by interchanging $\vec
x^{\ell-1}_{i_\ell}$ and $\vec x^{\ell-1}_{j_\ell}$ whenever $\vec
x^{\ell-1}_{i_\ell}>\vec x^{\ell-1}_{j_\ell}$.  The output of the
network for input $\vec x$ is $C(\vec x)=\vec x^k$, and
$\outputs(C)=\sset{C(\vec x)}{\vec x\in\{0,1\}^n}$.  The comparator
network $C$ is a \emph{sorting network} if all elements of
$\outputs(C)$ are sorted (in ascending order).
The zero-one principle~\cite{Knuth73} implies that a sorting
network also sorts 
sequences over any other totally ordered set, e.g.~integers.
\begin{wrapfigure}[3]{r}{0.18\textwidth}
\vspace*{-5ex}
  \raisebox{-2ex}{\includegraphics{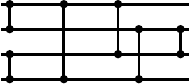}}
\end{wrapfigure}
The image on the right depicts a sorting network on 4
channels, consisting of 6 comparators.  The channels are
indicated as horizontal lines (with channel $4$ at the bottom),
comparators are indicated as vertical lines connecting a pair of
channels, and input values propagate from left to
right. The sequence of comparators associated with a picture
representation is obtained by a left-to-right, top-down
traversal. For example, the network depicted above is $(1,2); (3,4); (1,4); (1,3); (2,4); (2,3)$.

The optimal-size sorting network problem is about finding the smallest
size, $S(n)$, of a sorting network on $n$ channels.  In~1964,
Floyd and Knuth presented sorting networks of optimal size for $n\leq 8$
and proved their optimality~\cite{Knuth66}.
For nearly fifty years there was no further progress on this problem, until we established
that $S(9) = 25$~\cite{ourICTAIpaper} and, consequently, using a theoretical result on lower bounds~\cite{voorhis72}, that $S(10) = 29$.
Currently, the best known bounds for $S(n)$ are:
\begin{center}
\begin{tabular}{l|c|c|c|c|c|c|c|c|c|c|c|c|c|c|c|c}
$n$ & 1 & 2 & 3 & 4 & 5 & 6  & 7  & 8  & 9  & 10 & 11 & 12 & 13 & 14 & 15 & 16
\\ \hline
upper bound for $S(n)$ & 0 & 1 & 3 & 5 & 9 & 12 & 16 & 19 & 25 & 29 & 35 & 39 & 45 & 51 & 56 & 60\\
lower bound for $S(n)$\,&\,0\,&\,1\,&\,3\,&\,5\,&\,9\,&\,12\,&\,16\,&\,19\,&\,25\,&\,29\,&\,33\,&\,37\,&\,41\,&\,45\,&\,49\,&\,53\,
\end{tabular}
\end{center}

Our proof relies on a program that checks that there is no sorting network on $9$ channels with only $24$ comparators.
The algorithm exploits symmetries in comparator networks, in particular the notion of \emph{subsumption}.
Given two comparator networks on $n$ channels $C_a$ and $C_b$ and a permutation $\pi$ on $\{1,\ldots,n\}$,
we say that $C_a$ subsumes $C_b$ by $\pi$, and write $C_a\leq_\pi C_b$,
if there exists a permutation $\pi$ such that $\pi(\outputs(C_a)) \subseteq \outputs(C_b)$.
We will write simply $C_a\preceq C_b$ to denote that $C_a\leq_\pi C_b$ for some $\pi$.

Subsumption is a powerful mechanism for reducing candidate sequences of comparators when looking for sorting networks:
if $C_a$ and $C_b$ have the same size, $C_a\preceq C_b$ and there is a sorting network $C_b;C$ of size~$k$, then there also is a sorting network $C_a;C'$ of size~$k$.
This motivated the \emph{generate-and-prune} approach to the optimal-size sorting network problem: starting with the empty network, alternately add one comparator in all possible ways and reduce the result by
eliminating subsumptions.
More precisely, the algorithm
iteratively builds two sets $R^n_k$
and $N^n_k$ of $n$ channel networks of size $k$.
First, it initializes $R^n_0$ to contain only the empty comparator network. Then, it repeatedly applies two
types of steps, \algorithm{Generate} and \algorithm{Prune}.%

\begin{enumerate}
\item\algorithm{Generate}: Given $R^n_k$, construct $N^n_{k+1}$ by adding one comparator to each
  element of $R^n_k$ in all possible ways.
\item\algorithm{Prune}: Given $N^n_{k+1}$, construct
  $R^n_{k+1}$ such that every element of $N^n_{k+1}$ is subsumed by an element of $R^n_{k+1}$.
\end{enumerate}
The algorithm stops when a sorting network is found, in which case $\left|R^n_k\right|=1$.

Soundness of the algorithm relies on the fact that $N^n_k$ (and $R^n_k$) are \emph{complete} for the optimal size sorting
network problem on $n$ channels: if
there exists an optimal size sorting network on $n$ channels, then there exists one
of the form $C;C'$ for some $C\in N^n_k$ (or $C\in R^n_k$), for every $k$.

Computationally, the big bottleneck is the pruning step, where to find subsumptions we test all pairs of networks by looking at $9!\approx 3.6\times 10^5$ permutations -- and at the peak the set $N^9_k$ contains around $1.8\times10^7$ networks, so there are potentially $3.2\times10^{14}$ tests.
By extending generate-and-prune with the optimizations and extensive parallelization described in~\cite{ourICTAIpaper}, we were able to show that $S(9)=25$ in around three weeks of computation on $288$~threads.

However, the same optimizations that made the program work made it less trustworthy.
Therefore, we formalized the soundness of generate-and-prune in the theorem prover Coq with the goal of extracting a provenly correct checker of the same result~\cite{ourformalization} to Haskell.\footnote{The choice of Haskell as target language is pragmatic: preliminary experiments suggested that it was the fastest one for this project.}
In order to eliminate the search step in \verb+Prune+, this formalization is parameterized on an oracle, which produces triples $\langle C_a,C_b,\pi\rangle$ such that $C_a\leq_\pi C_b$.
This oracle is untrusted, so the checker will validate this subsumption and discard it if it cannot do so; but using it allows us to remove all search, while simultaneously making the number of tests linear in $N_n^k$, rather than quadratic.
It is implemented by reading the logs produced by the original execution of generate-and-prune, in which all successful subsumptions were recorded.
They amount to a total of $27$~GB, making this one of the largest computer-generated proofs ever.

The formalization defines \verb+comparator+ to be a pair of natural numbers and the type \verb+CN+ of comparator networks to be \verb+list comparator+.
We then specify what it means for a comparator network to be a sorting network on $n$ channels, and show that
this is a decidable predicate.
The details of the formalization of the theory of sorting networks can be found in~\cite{ourformalization}.

The implementation of generate-and-prune proceeds in several steps.
We translate \verb+Generate+ directly into Coq code, which we omit since it is straightforward and we will not discuss ir further.
As for \verb+Prune+, we closely follow the original pseudo-code in~\cite{ourICTAIpaper}.\footnote{Throughout this presentation we will always show transcribed Coq code, which is almost completely computational and preserved by extraction.}
{\small\begin{verbatim}
Definition Oracle := list (CN * CN * (list nat)).

Function Prune (O:Oracle) (R:list CN) (n:nat)
               {measure length R} : list CN := match O with
  | nil => R
  | cons (C,C',pi) O' => match (CN_eq_dec C C') with
      | left _ => R
      | right _ => match (In_dec CN_eq_dec C R) with
          | right _ => R
          | left _ => match (pre_permutation_dec n pi) with
              | right _ => R
              | left A => match (subsumption_dec n C C' pi' Hpi) with
                  | right _ => R
                  | left _ => Prune O' (remove CN_eq_dec C' R) n
end end end end end.
\end{verbatim}}
\verb+Prune+ processes each subsumption $\langle C,C',\pi\rangle$ given by the oracle sequentially and makes all the relevant checks: that $C\neq C'$ (\verb+left+ extracts as \verb+True+, \verb+right+ as \verb+False+), that $C\in R$, that $\pi$ represents a valid permutation, and that $C\leq_\pi C'$.  If all checks succeed, $C'$ is removed from $R$, otherwise the subsumption is discarded.
For legibility, we write \verb+pi'+ for the translation of $\pi$ into our representation of permutations, and \verb+Hpi+ for the proof term needed for the subsumption test.

Both \verb+Generate+ and \verb+Prune+ are proven to take complete sets of filters into complete sets of filters, as well as to satisfy some aditional properties necessary for the soundness of the algorithm.
These functions are then incorporated in a larger loop that applies them alternately.
The code uses \verb+OGenerate+, an optimized version of \verb+Generate+ that removes some networks using known results about redundant comparators that were implemented in the original algorithm and that are easily shown to be sound~\cite{ourICTAIpaper,Knuth73}.
This loop receives as inputs the number of channels $m$ and the number of iterations $n$, and returns an answer: \verb+(yes m k)+ if a sorting network of size $k$ was found; \verb+(no m k R)+ if a set \verb+R+ of comparator networks of size $k$ is constructed that is complete and contains no sorting network; or \verb+maybe+ if an error occurs.
The answer \verb+no+ contains some extra proof terms necessary for the correctness proof.
These are removed in the extracted checker, and since they make the code quite complex to read, we replace them by \verb+_+ below.

{\small\begin{verbatim}
Fixpoint Generate_and_Prune (m n:nat) (O:list Oracle) : Answer :=
  match n with
  | 0 => match m with
         | 0 => yes 0 0
         | 1 => yes 1 0
         | _ => no m 0 (nil :: nil) _ _ _
         end
  | S k => match O with
      | nil => maybe
      | X::O' => let GP := (Generate_and_Prune m k O') in match GP with
           | maybe => maybe
           | yes p q => yes p q
           | no p q R _ _ _ => let GP' := Prune X (OGenerate R p) p in
                               match (exists_SN_dec p GP' _) with
                                     | left _ => yes p (S q)
                                     | right _ => no p (S q) GP' _ _ _
end end end end.
\end{verbatim}}
Here \verb+Answer+ is the suitably defined inductive type of answers.
The elimination over \verb+exists_SN_dec+ uses the fact that we can decide whether a set contains a sorting network.
Correctness of the result is shown in the two theorems below. %
In these, the oracle \verb+O+ is universally quantified, reflecting that they hold regardless of whether the oracle is giving right or wrong information.
{\small\begin{verbatim}
Theorem GP_yes : forall m n O k, Generate_and_Prune m n O = yes m k ->
                (forall C, sorting_network m C -> length C >= k) /\
                 exists C, sorting_network m C /\ length C = k.

Theorem GP_no : forall m n O R HR0 HR1 HR2,
                     Generate_and_Prune m n O = no m n R HR0 HR1 HR2 ->
                     forall C, sorting_network m C -> length C > n.
\end{verbatim}}

The extracted code for \verb+Generate_and_Prune+ is a function that takes two natural numbers $m$ and $n$ and a list of oracles, applies generate-and-prune on $m$ channels for $n$ iterations using the oracles, and returns \verb+yes m k+ or \verb+no m k R+.
The soundness theorems guarantee that these answers have a mathematical meaning.

\section{Reducing runtime of the pruning step}
\label{sec:runtime1}

\begin{figure}
\centering
\resizebox{\textwidth}{!}{\input{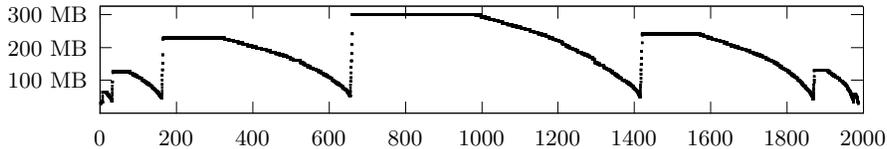}}
\caption{Memory usage (MB/min) during the verification of the proof for $8$~channels.}
\label{fig:mem}
\end{figure}

Figure~\ref{fig:mem} displays the memory usage during the validation of the proof for $8$~channels.
The exact values are immaterial, but we can very easily trace the execution of the algorithm by noting that every upwards jump corresponds to \verb+Generate+, whereas the descending curve corresponds to \verb+Prune+.
The picture shows that the three most costly iterations account for almost $90\%$ of the execution time.
For $9$~channels, there are four costly iterations, and the imbalance will be even greater, as the differences in size between the sets $N^9_k$ are much more significant.

The biggest cost in the execution of the checker is in the pruning step, as was already the case with the original, uncertified program.
The use of the oracle allows us to bypass the original search, but the algorithm is still very inefficient: for every subsumption, it iterates through the set being pruned to verify that the subsuming network is there and to remove it.
Due to lazy evaluation in Haskell, these verifications are made in a single pass; but execution time is still quadratic on the number of generated networks.

In this section we take advantage of the offline nature of our oracle, and show that we can greatly improve the algorithm using the fact that we already know all the subsumptions we will make.
Indeed, we need to do three things.
\begin{enumerate}
\item Check that all subsumptions are valid.
\item Remove all subsumed networks.
\item Check that all networks used in subsumptions are kept.
\end{enumerate}

Each subsumption in step~1 is checked individually, so this step scales linearly in the number of networks.
The other two steps can be significantly improved.

\subsection{Optimizing the removal step}

In theory, step~2 could be substantially optimized by delaying the removals until all subsumptions have been read: if we obtained the networks to be removed from the oracle in the same order as we generate them in the checker, then we could remove all subsumed networks with one single pass over the whole set, instead of having to iterate through the set of networks for each subsumed network.

This is the first time that the symbiosis between the prune algorithm and the implementation of the untrusted oracle becomes a key ingredient for optimization. As we use an \emph{offline} oracle~\cite{ourformalization}, we can actually reorder the oracle information to suit the needs of the checker with an efficient (untrusted) pre-processor. An inspection of the definition of \verb+Generate+ shows that comparators are added in lexicographic order, and we can pre-process the oracle information such that the subsumptions are provided in the same order.
Then we can define a function \verb+remove_all+ to complete step~2 in linear time by simultaneously traversing the list of subsumed networks and the list of all networks and removing all elements of the former from the latter.

\subsection{Optimizing the presence check}

Unfortunately, one cannot do a similar optimization to step~3 immediately, since sorting the oracle information by the subsumed networks will yield an unsorted sequence of subsuming networks.
However, we can proceed in a different way: rather than checking that the subsuming networks are kept at each step, only check that they are present in the final (reduced) set.
This will still be a quadratic algorithm, but relative to the size of the final set -- which, in the most time-consuming steps, is only around $5\%$ of the size of the original one.

This idea again requires an important change to the oracle implementation, this time in the subsumptions presented by the oracle.
As it happens, there are often chains of subsumptions $C_1\preceq C_2\preceq\ldots\preceq C_n$, which pose no problem for the original algorithm, but would result in a false negative result of the checker, if we were to check the presence of the subsuming networks in the final set. Consider e.g.\ $C_2$, which is used to remove $C_3$, but which is itself removed by $C_1$.

However, we can benefit from the offline character of the oracle and use the transitivity of subsumption to transform such chains of subsumptions into ``reduced'' subsumptions $C_1\preceq C_2$, $C_1\preceq C_3$, \ldots, $C_1\preceq C_n$.
This again requires pre-processing the oracle information, identifying such chains and computing adequate permutations for the new resulting subsumptions.

In order to achieve this, we implemented a data structure in the pre-processor that we term a subsumption graph: a labeled directed graph whose nodes are comparator networks, and where there is a edge from $C'$ to $C$ labeled by $\pi$ if $C \leq_\pi C'$.
Once we have built the full graph for one pruning step, we can obtain the reduced oracle information as follows: (i) find all non-empty paths in the graph ending in a node without outgoing edges; (ii) starting with the identity permutation, traverse each such path while composing the permutations on the edges; (iii) the start- and end-node of each path, together with the resulting permutation, describe one reduced subsumption.
The oracle then provides the reduced subsumptions instead of the original ones.

The formalized definitions for the improved pruning step now look as follows.
Functions \verb+oracle_ok_1+ and \verb+oracle_ok_2+ perform steps~1 and~3 above, and \verb+Prune+ uses \verb+remove_all+ to perform step~2.
{\small\begin{verbatim}
Fixpoint oracle_ok_1 (n:nat) (O:Oracle) : bool := match O with
  | nil => true
  | (C,C',pi) :: O' => match (pre_permutation_dec n pi) with
             | right _ => false
             | left A => match (subsumption_dec n C C' pi' Hpi) with
                    | right _ => false
                    | left _ => oracle_ok_1 n O'
end end end.

Fixpoint oracle_ok_2 (O:Oracle) (R:list CN) : bool := match O with
  | nil => true
  | (C,_,_)::O' => match (In_dec CN_eq_dec C R) with
                   | left _ => oracle_ok_2 O' R
                   | right _ => false
end end.

Definition Prune (O:Oracle) (R:list CN) (n:nat) : list CN :=
  match (oracle_ok_1 n O) with
  | false => R
  | true => let R' := remove_all CN_eq_dec (map snd (map fst O)) R in
            match (oracle_ok_2 O R') with
            | false => R
            | true => R'
end end.
\end{verbatim}}

This approach is completely modular: after we reprove the lemmas regarding the correctness of \verb+Prune+, the proofs for the whole algorithm mostly go through unchanged, and where tweaking of the proofs is necessary, the changes are trivial and require no deep insights into the proofs.

\subsection{Practical impact on runtime}
\begin{figure}[b]
\centering
\resizebox{\textwidth}{!}{\input{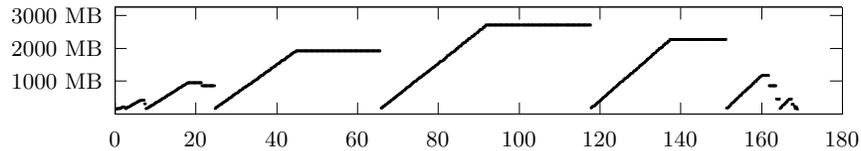}}
\caption{Memory usage (MB/min) verifying the proof for $8$~channels, after optimizations.}
\label{fig:memquick}
\end{figure}

In the following table, we compare the runtime of the original implementaton of the proof checker with the improved one presented in this section. We focus on the case of $8$ inputs, the largest case that we can systematically handle. 
\begin{center}
\begin{tabular}{l|c|c}
configuration & original algorithm & improved algorithm\\ \hline
runtime & $1985m$ & $167m$
\end{tabular}
\end{center}
Clearly, we see an order of magnitude improvement for $8$ inputs. We also ran the first $10$ pruning steps of $9$ and infer an even larger improvement for $9$ inputs, bringing down the expected runtime from two decades to several months.
The much lower weight of \verb+Prune+ is patent in the new memory trace (Figure~\ref{fig:memquick}).

\section{Reducing memory footprint by tuning the extraction}
\label{sec:memory1}

The contributions of the previous section left us with a checker that was nearly fast enough, but that had too large memory requirements due to reading all subsumptions at once, rather than processing them one by one. Our attempts to run the checker for $9$ inputs quickly drained the %
available computing resources, and we estimated that more than $200$~GB of RAM would be needed.
Profiling showed that most of the memory was being taken up by lists and natural numbers -- not surprising, since the checker is producing millions of comparator networks.
But when, at the peak, we potentially need to store $18$~million networks$\times 15$~comparators$\times 2$~channels, the Peano representation of natural numbers in Coq is extremely expensive, even with all numbers ranging from $0$ to $8$.

The most natural idea was to extract natural numbers to Haskell native types.
In general, this loses the guaranteed correctness of the extracted program; but in this particular example it not pose significant risks, as natural numbers are identifiers for channels and not objects with which to do computations.
This means that only five Haskell functions are needed: \verb+succ+, \verb+(=)+, \verb+(<)+, \verb+(-)+ and \verb+max+, besides the recursor
\begin{verbatim}
(\ fO fS n -> if n==0 then (fO __) else fS (n-1))
\end{verbatim}
(Function \verb+max+ is used only in the definition of predecessor, while \verb+{-}+ is used only in the recursor.)
Furthermore, they only operate on the numbers $0$ to $8$ (except for \verb+succ+, which goes up to $25$), so it is easy to verify exhaustively that they are correct.
As a side-effect, we also need to extract booleans to the native \verb+Bool+ type (which has exactly the same definition as extracting from the Coq type), and since we do not use any functions on \verb+Bool+ this is also not a problem.

\subsection{Practical impact on memory usage}
In the following table, we compare the memory usage of the extracted Peano numerals (an algebraic data structure with constructors \verb+0+ and \verb+S+) against several native representations. Once again, we consider the case of $8$ inputs.
\begin{center}
\begin{tabular}{l|c|c|c|c}
representation & Peano naturals & $64$-bit integers & $8$-bit integers & enum\\ \hline
memory usage (MB) & 2536 & 844 & 1669 & 999\\
\end{tabular}
\end{center}
We clearly see that native $64$-bit integers (\verb+int+) take significantly less memory than %
Peano numerals.
Interestingly, other datatypes perform worse: although \verb+enum+ or \verb+int8+ in general use less memory than \verb+int+, the fact that the Haskell compiler keeps a small store of ``reusable'' integers in the heap actually makes \verb+int+ perform better, memory-wise, than either \verb+int8+ or \verb+enum+.

We also experimented using Haskell lists instead of extracted Coq lists, but this does not help: these datatypes are isomorphic, and we still use a recursor instead of pattern-matching.

\section{Optimizing data structures}
\label{sec:runtime2}

With all these optimizations in place, the task of checking the full proof for $9$~inputs became just beyond reachable.
Experiments that the memory consumption for each iteration of generate-and-prune was linear on the number of comparators in the subsumptions in the oracle; this allowed us to estimate the total memory required at $80-90$~GB.
Likewise, the execution times for the first $12$ steps showed a linear dependency on the total number of generated nets (which seemed reasonable and hard to improve, since we need to generate them explicitly and then prune them) and a quadratic dependency on the size of the pruned set (due to the check that all networks used in subsumptions are kept).
A rough estimate based on a least squares fit of the data yielded around four months for the whole execution.

We therefore focused on more localized aspects of the formalization in order to bring these requirements down and actually verify the complete proof.
Our decision on what constitutes ``reasonable'' is directly related to the available resources: $500$~hours of computation on a computer with $64$~GB of RAM memory. In this section we focus on runtime.

\subsection{Using binary search trees to decide membership}

The step that we felt was most inefficient was the verification in \verb+oracle_ok_2+, where we iterate over all networks used in subsumptions and check that they occur in the pruned set.

There are two reasons why the implementation of this step is not satisfactory.
First, since we are iterating over all subsumptions, we repeatedly test the same network many times: at peak, there are about $20$ times as many subsumptions as networks in the pruned set.
Secondly, these subsumptions are unordered, but the list of pruned networks is ordered; however, since Coq lists do not have direct access, we are still forced to look for them in linear time.
This means that this step takes time proportional to both the number of subsumptions and the number of pruned networks, and is thus roughly quadractic on the latter.

Ideally, we would like to do something similar to the optimization of the pruning step itself, where by ensuring the list of all networks and the list of networks to be removed are ordered in the same way we can solve the problem in linear time.
However, the trick we used before is no longer applicable, since we cannot change the order of the oracle.

Instead, we pursued the idea of sorting the networks used in subsumptions (and removing duplicates as we do so).
In order to do this efficiently, we changed the data structure storing these networks, from a list to a search tree.
This required enriching our formalization with a type of binary trees and operations for adding and retrieving the minimum element of such a tree.

In keep with the remainder of the formalization~\cite{ourformalization}, we defined binary trees without any restrictions, together with a predicate stating that a binary tree is a search tree.
This is similar to the formalization of binary trees in~Chapter~11 of~\cite{CoqArt}; however, that formalization only considered trees over Coq integers, whereas we formalize binary trees over an arbitrary type \verb+T+ over which we have a comparison function.

{\small\begin{verbatim}
Inductive BinaryTree (T:Type) : Type :=
  nought : BinaryTree
  | node : Tree -> BinaryTree -> BinaryTree -> BinaryTree.
\end{verbatim}}

We then define predicates \verb+BT_in+ to test that an element occurs in a binary tree, \verb+BT_wf+ to check that a binary tree is a search tree,
and the usual function \verb+BT_add+ to add an element to a tree.
For efficiency, we also define a function \verb+BT_split+ that simultaneously computes the minimum element of a search tree \emph{and} the tree obtained by removing it.

{\small\begin{verbatim}
Fixpoint BT_split (T:Type) (BT:BinaryTree T) (val:T) : T * BinaryTree :=
  match BT with
  | nought => (val,nought)
  | node t nought R => (t,R)
  | node t L R => let (t',L') := BT_split L val in (t',node t L' R)
  end.
\end{verbatim}}
We show that the functions defined work correctly on search trees; in particular, any object of type \verb+BinaryTree+ built from \verb+nought+ by repeated application of \verb+BT_add+ satisfies \verb+BT_wf+.
Then, we changed the implementation of \verb+oracle_ok_1+ to return also a binary tree, proved that this is a search tree containing all networks used in the subsumptions given by the oracle,
and rewrote \verb+oracle_ok_2+ to run in only slightly superlinear time.

{\small\begin{verbatim}
Fixpoint oracle_ok_2 (BT:BinaryTree CN) (R:list CN) := match BT,R with
  | nought, _ => true
  | _, nil => false
  | _, C' :: R' => let (C,BT') := (BT_split BT nil) in
                   match (OCN_eq_dec C C') with
                   | left _ => oracle_test BT' R'
                   | right _ => oracle_test BT R'
end end.
\end{verbatim}}

Some of the proofs in the pruning step required a bit of adaptation, since they now rely on lemmas over \verb+BinaryTree+s instead of \verb+list+s, but the changes were localized to this part of the formalization.

The recursive call in \verb+oracle_ok_2+ is on the remainder of the list, so the total execution time depends on the length of this list and the depth of the search tree \verb+BT+.
Before experimenting with the newly extracted program, we exhaustively ran the oracle sources through a small Java program to check how balanced the constructed search trees would be.
The maximum depth is only $94$ (corresponding to a very unbalanced tree, but much better than the previous list), and for the two biggest sets of subsumptions we actually obtain trees of depth $69$, storing $848{,}914$ networks, in one case, or $568{,}287$, in the other.

\subsection{Using binary search trees for subsumption checking}

The availability of binary trees unexpectedly opened the door to another improvement in the program: the subsumption test itself.
Lemma \verb+subsumption_dec+ states that $C\leq_\pi C'$ is decidable, and the proof simply proceeds by computing $\outputs(C)$ and $\outputs(C')$ and directly checking that $\pi(\outputs(C))\subseteq\outputs(C')$.
Since the number of outputs is fixed, this check takes almost constant time (computing the outputs becomes slightly more time-consuming as the networks grow bigger, but this is not noticeable), but on $9$ channels the lists of outputs contain $512$ elements, and again they have many repetitions and are reasonably unordered.

Therefore, we experimented with reproving \verb+subsumption_dec+ by storing the computed outputs in a search tree rather than in a list.
The impact on performance was stunning: since the execution time was now dominated by the validation of all the subsumptions, we were able to check the proof for $8$ inputs in less than half the time.

\subsection{Practical impact on runtime}
The following table summarizes the impact of the contributions in this section on the verification of the proof for $8$~inputs.
\begin{center}
\begin{tabular}{l|c|c|c}
configuration & original & tree-based presence check & everything tree-based\\ \hline
runtime & $126m$ & $111m$ & $48m$ 
\end{tabular}
\end{center}
Using trees for checking for the presence of subsuming networks has a moderate impact on $8$ inputs.
However, this impact becomes greater as the number of inputs grows: experiments with the initial pruning steps for $9$ inputs gave an estimated runtime reduction of $30\%$. Experiments suggest that using both optimizations yields approx.\ $70\%$ reduction of runtime on $9$ inputs.

One might wonder whether we could not use search trees in the original formalization and gain a similar speedup.
The answer is negative: the improvement stems both from the numerous repetitions among the subsuming networks and from their failure to be ordered.
The generation step produces networks that are both ordered and without repetitions, whence the result of storing them in a search tree would be isomorphic to a list.

\section{G\"odelizing comparators to reduce memory footprint}
\label{sec:memory2}
At this point, the remaining bottleneck was memory, and we again shifted focus from runtime to reducing the memory footprint.
We decided to take advantage of Haskell's caching of small integers by using a G\"{o}delization of comparators: represent each comparator (a pair of natural numbers) by a single natural number, using the bijection $\varphi(i,j)=\frac12j\times(j-1)+i$.
This happens to map very nicely to the function \verb+all_st_comps+ described earlier, since the comparator $(i,j)$ is exactly the $\varphi(i,j)$-th element of \verb+all_st_comps n+ (as long as $i,j<n$).

We then defined a type \verb+OCN := list nat+ of optimized comparator networks and a mapping to \verb+CN+.
Using this mapping, it was possible to reimplement \verb+Generate+ and \verb+Prune+ to run on lists of \verb+OCN+, while reusing all the old theory about comparator networks.
From a formalization point of view, it was also the most reasonable option, as it keeps a consistent theory of comparator networks formalized according to intuition, and uses a more efficient representation only for implementation purposes.

The following table compares memory usage of representing comparators by a pair of \verb+int+ or by one G\"{o}delized \verb+int+, for the case of $8$ inputs. 
\begin{center}
\begin{tabular}{l|c|c}
comparators & explicit & G\"{o}delized\\ \hline
memory usage (MB) & $844$ & $541$ 
\end{tabular}
\end{center}
Assymptotically, this change reduces memory consumption to just over one half: for each comparator we are now just storing one number instead of a pair of numbers. Again, experiments suggest that the improvement for $9$ inputs is greater than for the case detailed.
There is some overhead of mapping from \verb+CN+ to \verb+OCN+ to test subsumptions, but it is offset by an improvement in pruning times due to testing for equality directly on \verb+OCN+.

\bigskip

With all these optimizations in place, our checker was able to verify the original proof of optimality of $25$ comparators for sorting $9$ inputs, using the available proof witnesses.
The verification took $163.8$~hours, or just under one week, required a maximum of $50.05$~GB of RAM, and returned the answer \verb+yes 9 25+.

\section{Conclusion}
\label{sec:conclusion}

The contributions of Sections~\ref{sec:runtime1}--\ref{sec:memory2} allowed us to run a formal validation of the proof from~\cite{ourICTAIpaper} that $25$ comparators suffice for sorting $9$ inputs, using the formalization of the theory of sorting networks described in~\cite{ourformalization}.%

We also showed that it is feasible to optimize extracted code without significantly changing the underlying formalized theory, and therefore the latter can be developed without excessive concerns about the extracted code.
Indeed, the original formalization closely follows Knuth~\cite{Knuth73}, with the new theoretical results from~\cite{ourICTAIpaper} and a straightforward implementation of the algorithm therein proposed.
While this theory took three months to formalize, each of the changes described in this paper required only around one day, as they amounted to changing localized parts of the checker and reproving their properties.
In other words, the optimizations were obtained by concentrating on the computational aspects of the checker
without needing to worry about the underlying theory.

These results support our choice of 
an \emph{offline} untrusted oracle for the original formalization~\cite{ourformalization}
as it allows for a nice separation between the development of the theory and the optimization of the checker, as well as giving us the possibily of exploring the interplay between the checker and oracle.

We plan to test %
this approach %
to validate other search-intensive, large-scale computer-generated proofs.

\section*{Acknowledgements}

We would like to thank Pierre Letouzey for suggesting and helping with extracting to Haskell native types, S\o ren Haagerup for helping with profiling, and Michael Codish for his support and his enthusiasm about sorting networks.

The authors were supported by the Danish Council for Independent Research, Natural Sciences. Computational resources were generously provided by the Danish Center for Scientific Computing.

\bibliographystyle{plain}
\bibliography{extraction}

\end{document}